\documentclass[aps,prl,superscriptaddress,reprint,amsfonts,amssymb,showpacs]{revtex4-1}
\pdfoutput=1
\usepackage{graphics}
\usepackage{hyperref}

\begin{document}

\title{Normal-mode splitting in the coupled system of hybridized nuclear magnons and microwave photons}

\author{L.V. Abdurakhimov}
\email{leonid.abdurakhimov@oist.jp}
\affiliation{Okinawa Institute of Science and Technology (OIST) Graduate University, Onna, Okinawa 904-0495, Japan}
\author{Yu.M. Bunkov}
\affiliation{Institute Neel, CNRS, Grenoble, 30842, France}
\author{D. Konstantinov}
\affiliation{Okinawa Institute of Science and Technology (OIST) Graduate University, Onna, Okinawa 904-0495, Japan}

\date{\today}

\begin{abstract}
In the weak ferromagnetic MnCO$_3$ system, a low-frequency collective spin excitation (magnon) is the hybridized oscillation of nuclear and electron spins coupled through the hyperfine interaction.  By using a split-ring resonator, we performed transmission spectroscopy measurements of MnCO$_3$ system and observed, for the first time, avoided crossing between the hybridized nuclear magnon mode and the resonator mode in the NMR-frequency range. The splitting strength is quite large due to the large spin density of $^{55}$Mn, and the cooperativity value $C=0.2$ (magnon-photon coupling parameter) is close to the conditions of strong coupling. The results reveal a new class of spin systems, in which the coupling between nuclear spins and photons is mediated by electron spins via the hyperfine interaction, and in which the similar normal-mode splitting of the hybridized nuclear magnon mode and the resonator mode can be observed.
\end{abstract}


\maketitle

\paragraph{Introduction.}

When two oscillators are coupled, their otherwise degenerate modes split into two non-crossing hybridized spectrum branches. The phenomenon of avoided crossing is ubiquitous and have been observed in many systems ranging from a classical coupled spring-mass system~\cite{novotny2010} to interacting atoms and photons in an optical cavity~\cite{zhu1990, kimble1992}, to paramagnetic spin ensembles coupled to microwave cavities~\cite{chiorescu2010,schuster2010,kubo2010,abe2011,probst2013,eddins2014}. Systems with strong coupling of microwave photons and spin ensembles are of particular interest since they are proposed to be important building elements of hybrid quantum computers~\cite{wu2010,duty2010}. Recently, strong coupling between ferromagnetic electron-spin magnons and microwave photons has been observed in yttrium iron garnet (YIG, Y$_3$Fe$_5$O$_{12}$)~\cite{huebl2013, zhang2014, tabuchi2014,tabuchi2014_2, goryachev2014, bhoi2014}. In comparison with the paramagnetic diluted spin ensembles, ferromagnetic materials provide a much higher density of spins which are ordered due to the exchange interaction. Therefore, strong coupling in ferromagnetic systems is characterized by large values of the coupling strength since the coupling constant between a microwave cavity mode and a collective spin excitation (magnon) $g_m$ is proportional to the square root of the total number of polarized spins $N$~\cite{imamoglu2009, agarwal1984}, i.e., $g_m=g_0 \sqrt{N}$, where $g_0$ is the coupling strength of a single spin to the cavity.

Despite the significant progress in studies of strong coupling between microwave cavity mode and electron spin systems, realization of strong coupling between photons and a nuclear spin ensemble is still a challenging problem. Paramagnetic nuclear spins are weakly polarized, and nuclear magnetic moments are typically about $10^3$ times smaller than electron magnetic moments. Therefore, the coupling strength between nuclear spins and photons is much weaker in comparison with the electron-spin-photon coupling. To the best of our knowledge, nobody has reported the coupling of a nuclear spin ensemble to a microwave cavity. On the other hand, quantum techniques involving nuclear spins are of great interest due to long nuclear spin decoherence times~\cite{brown2011}.

In this Letter, we report the first observation of a significant coupling between a nuclear spin ensemble in MnCO$_3$ system and microwave photons in a cavity. It is manifested as the normal-mode splitting of two coupled oscillators: the hybridized nuclear magnon --- homogeneous coupled-spin precession of the nuclei and the electrons (spin wave with $k=0$) --- and the ac electromagnetic field in the microwave resonator.

\begin{figure}
\includegraphics{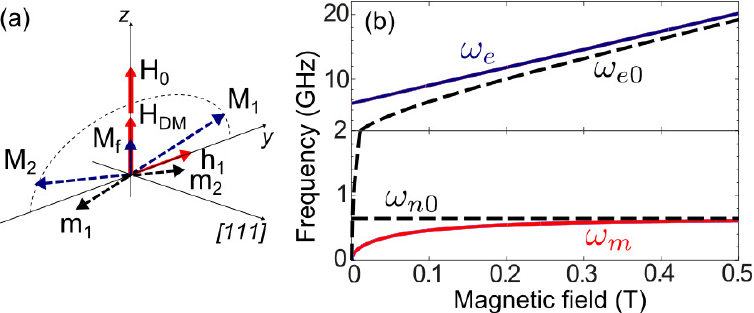}
\caption{\label{fig1} (color online) (a) Magnetic structure of antiferromagnetic MnCO$_3$ in the (111) plane ($y$-$z$ plane). Electron magnetic sublattices with magnetizations $M_1$ and $M_2$, respectively, are slightly canted due to the Dzyaloshinskii-Moriya interaction represented by the effective magnetic field $H_{DM}$.  As a result, the non-zero net magnetization $M_f$ causes weak ferromagnetism of MnCO$_3$. Nuclear magnetic sublattices with magnetizations $m_1$ and $m_2$, respectively, are coupled to the electron sublattices through the hyperfine interaction fields $\mathbf{H_n}=-A\mathbf{M_i} \: (i=1,2)$. $H_0$ is the external static magnetic field, $h_1$ is the microwave field. (b) Due to the hyperfine coupling, oscillation modes of the electron spin system $\omega_{e0}$ (dashed curve) and the nuclear spin system $\omega_{n0}$ (dashed horizontal line) split into two hybridized branches:  the nuclear-ferromagnetic mode $\omega_m$ (red curve) and the electron-ferromagnetic mode $\omega_e$ (blue curve). Curves are shown in the two subpanels with different scales along the vertical axis.}
\end{figure}

The magnetic structure of MnCO$_3$ crystals was extensively investigated previously~\cite{date1960, deGennes1963, turov1966, borovik-romanov1965, shaltiel1966, tulin1969, borovik-romanov1984, borovik-romanov1964}. The properties of MnCO$_3$ are summarized hereby briefly. MnCO$_3$ is a pale-pink transparent crystal with rhombohedral crystal structure. Below the N\'eel temperature $T_N \approx 32.5$\,K, $^{55}$Mn electron moments form two sublattices that lie in the (111) plane and are canted from perfectly antiparallel antiferromagnetic order due to the Dzyaloshinskii-Moriya (DM) spin-spin interaction (see FIG.~\ref{fig1}(a)).  Each electron spin sublattice is subjected to an exchange field $H_E$ of about 340\,kOe~\cite{shaltiel1966}, an effective magnetic field of DM interaction $H_{DM}$ equal to 4.4\,kOe~\cite{borovik-romanov1964, borovik-romanov1965, shaltiel1966, tulin1969, borovik-romanov1984}, and small anisotropy fields which are not relevant to our experiments. Thus, a canting angle $\alpha$ is about one degree, and MnCO$_3$ is a weak ferromagnet (other examples of weak ferromagnets include $\alpha$-Fe$_2$O$_3$~\cite{pincus1960, mazurenko2005}, Y$_2$CuO$_4$~\cite{rouco1994}, FeBO$_3$~\cite{dmitrienko2014}). 

Important feature of MnCO$_3$ magnetic structure is that $^{55}$Mn nuclear spins ($I=5/2$) are coupled to electron spins ($S=5/2$) due to the hyperfine interaction~\cite{shaltiel1966, tulin1969, borovik-romanov1984}(see FIG.~\ref{fig1}(a)). An effective magnetic field acting on the nuclear spin system from the electron spin system $\mathbf{H_n}$ is characterized by the hyperfine coupling constant $A$, $\mathbf{H_n}=-A \mathbf{M}$, where $\mathbf{M}$ is the magnetization of the electrons of that sublattice on which sites nuclear spins are located. In the magnetically ordered state below the N\'eel temperature, the magnetization $M$ is near its maximum value, and the field $H_n$ reaches tremendous values of about $6 \times 10^5$\,Oe. Similarly, nuclear spins affect electron spins through the hyperfine field $\mathbf{H_{en}}=-A \mathbf{m}$, where $\mathbf{m}$ is the magnetization of the nuclear sublattice. The hyperfine interaction causes hybridization of the oscillations of electron and nuclear spin systems. In the case of the canted antiferromagnet MnCO$_3$, the analysis of the linearized equations of spin motion gives four frequency modes, two for the electron system and two for the nuclear system~\cite{deGennes1963}. One of the electron-frequency modes corresponds to an antiferromagnetic resonance, while the other represents a modified electron-ferromagnetic resonance (EFMR) related to the weak ferromagnetism of MnCO$_3$. Correspondingly, there are two hybridized nuclear-frequency modes: a mode of the nuclear spins oscillation coupled to the electron antiferromagnetic mode, and a mode of the nuclear spins oscillation coupled to the electron ferromagnetic mode.

In the experiments reported here, the MnCO$_3$ system was driven in NMR frequency range by a small magnetic microwave field $\mathbf{h_1}$ that lied in the (111) plane and was perpendicular to the constant external magnetic field $\mathbf{H_0}$ lied at the same plane, i.e., $\mathbf{h_1} \perp [111] \perp \mathbf{H_0}  $. In this case, only the mode of nuclear spins oscillation coupled to the EFMR mode can be excited at the frequency given by the linear approximation formula~\cite{deGennes1963, turov1966, borovik-romanov1965, shaltiel1966, tulin1969}:
\begin{equation}
\label{eq1}
\omega_m^2 \equiv \omega_n^2 = \omega_{n0}^2 \left( 1-\frac{\omega_{EN}^2}{\omega_e^2}\right).
\end{equation}
Here, $\omega_n$ and $\omega_e = \sqrt{\omega_{e0}^2 + \omega_{EN}^2}$ are the hybridized NMR and EFMR frequencies, $\omega_{n0} = \gamma_n H_n$ and $\omega_{e0} = \gamma_e \sqrt{ H_0 (H_0+H_{DM})}$ are the unmixed NMR and EFMR frequencies, $\omega_{EN}=\gamma_e \sqrt{2 H_E A m_0}$ is the hyperfine interaction parameter, $\gamma_n\approx 2\pi \times 11$\,MHz/T is the $^{55}$Mn nuclear gyromagnetic ratio~\cite{stone2005}, $\gamma_e=2\pi \times 28$\,GHz/T is the electron gyromagnetic ratio, and $m_0 = \langle m \rangle $ is the equilibrium nuclear magnetization. The average nuclear magnetization $m_0$ is determined by the hyperfine field $H_n$ as $m_0 = \chi_n H_n$, with  the nuclear magnetic susceptibility $\chi_n$ changing with nuclear temperature $T_n$ according to Curie's law, $\chi_n \sim T_n^{-1}$.  Figure~\ref{fig1}(b) shows the nuclear-ferromagnetic mode frequency $\omega_m$ and the electron-ferromagnetic mode frequency $\omega_e$ calculated from the equations~(\ref{eq1}) and the related expressions for $T=1.15$\,K. We used the following values for the corresponding parameters: $\omega_{n0}/2 \pi = 640$\,MHz, $H_{DM}=4.4$\,kOe, $2 H_E A m_0=5.8\times10^6/T_n$\,(Oe$^2\times$K)$\approx 5\times10^6$\,Oe$^2$ (we put $T_n \approx T$). The values were taken from~\cite{borovik-romanov1964, borovik-romanov1965, shaltiel1966,tulin1969, borovik-romanov1984}. Thus, the hybridization of nuclear and electron spins leads to the shift of the NMR frequency  --- so called ``frequency pulling'' effect~\cite{deGennes1963} --- which has been also observed in CsMnF$_3$, FeBO$_3$, CoCO$_3$, and other materials with the strong hyperfine interaction~\cite{borovik-romanov1984}.

It is important to note that an effective microwave field $h_n$ acting on the nuclei is considerably larger than the applied microwave field $h_1$. Indeed, although the angle of rotation $\varphi$ of the net electron magnetization $M_f$ (and, correspondingly, each of magnetizations $M_1$ and $M_2$) is very small, $\varphi \approx (h_1/H_0)(\omega_{e0}^2/\omega_e^2)$~\cite{shaltiel1966}, each electron sublattice produces a large varying field at the nuclei through the hyperfine field $H_n$. Each nuclear sublattice sees the microwave field $h_n=\varphi H_n = h_1(H_n/H_0)(\omega_{e0}^2/\omega_e^2)$ which is enhanced by the factor $\eta=(H_n/H_0)(\omega_{e0}^2/\omega_e^2)\approx 150$ relative to the applied microwave field $h_1$~\cite{shaltiel1966,turov1966}. 

\begin{figure}
\includegraphics{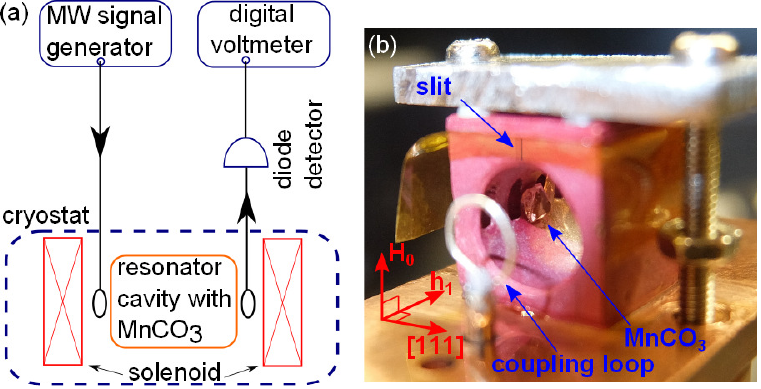}
\caption{\label{fig2} (color online) Experimental technique. (a)~Schematic diagram of the experimental apparatus for transmission spectroscopy. (b)~The sample of MnCO$_3$ was placed inside the split-ring resonator cavity in a such way that the constant magnetic field $\mathbf{H_0}$, the microwave magnetic field $\mathbf{h_1}$, and the [111] crystal axis were mutually perpendicular.}
\end{figure}

\paragraph{Experimental setup.}
In our experiments, we studied the coupling between the hybridized nuclear-electron spin system MnCO$_3$ and the microwave mode of a split-ring resonator. The experimental setup is shown schematically in FIG.~\ref{fig2}. We performed measurements of microwave power transmitted through the resonator as a function of probe microwave frequency and static magnetic field at the  temperature $T \approx 1.15$\,K. The split-ring resonator is shown in FIG.~\ref{fig2}(b), and, basically, can be approximated by a ring that is split by a gap~\cite{hardy1981}. The equivalent circuit of the split-ring resonator is the resonant LC circuit determined by the inductance of the single turn coil and the capacitance of the gap. In our experiments, the split-ring resonator was a copper cube with dimensions of 10$\times$10$\times$10\,mm$^3$, with a centered open-ended bore of about 7\,mm in diameter. The volume of the bore cavity $V_c$ was about 380\,mm$^3$. To form the gap of the split-ring resonator, a slit of 0.1\,mm was made along the middle line of the top resonator wall, parallelly to the axis of the bore. Thin film of Kapton with thickness of about 0.01\,mm was inserted between the cube face with the slit and an additional copper plate with dimensions of 10$\times$10$\times$1 mm$^3$, forming a capacitor for adjustment of resonant frequency. In our measurements, the resonant frequency of the split-ring resonator was about $\omega_c/2 \pi \approx 594$\,MHz, and Q-factor was about $Q_c=100$ which is determined by dielectric losses in the Kapton film.

The MnCO$_3$ single-crystal sample was in the form of a rhombus-shaped plate about 0.7\,mm thick, with diagonals of approximately 2.4\,mm and 2.7\,mm. The [111] crystal axis was perpendicular to the plane of the plate. The mass of the sample was about 8\,mg that corresponds to the number of $^{55}$Mn atoms of about $N=4\times10^{19}$. The sample was glued by GE Varnish to the wall inside the cavity of the split-ring resonator.

\begin{figure*}
\includegraphics{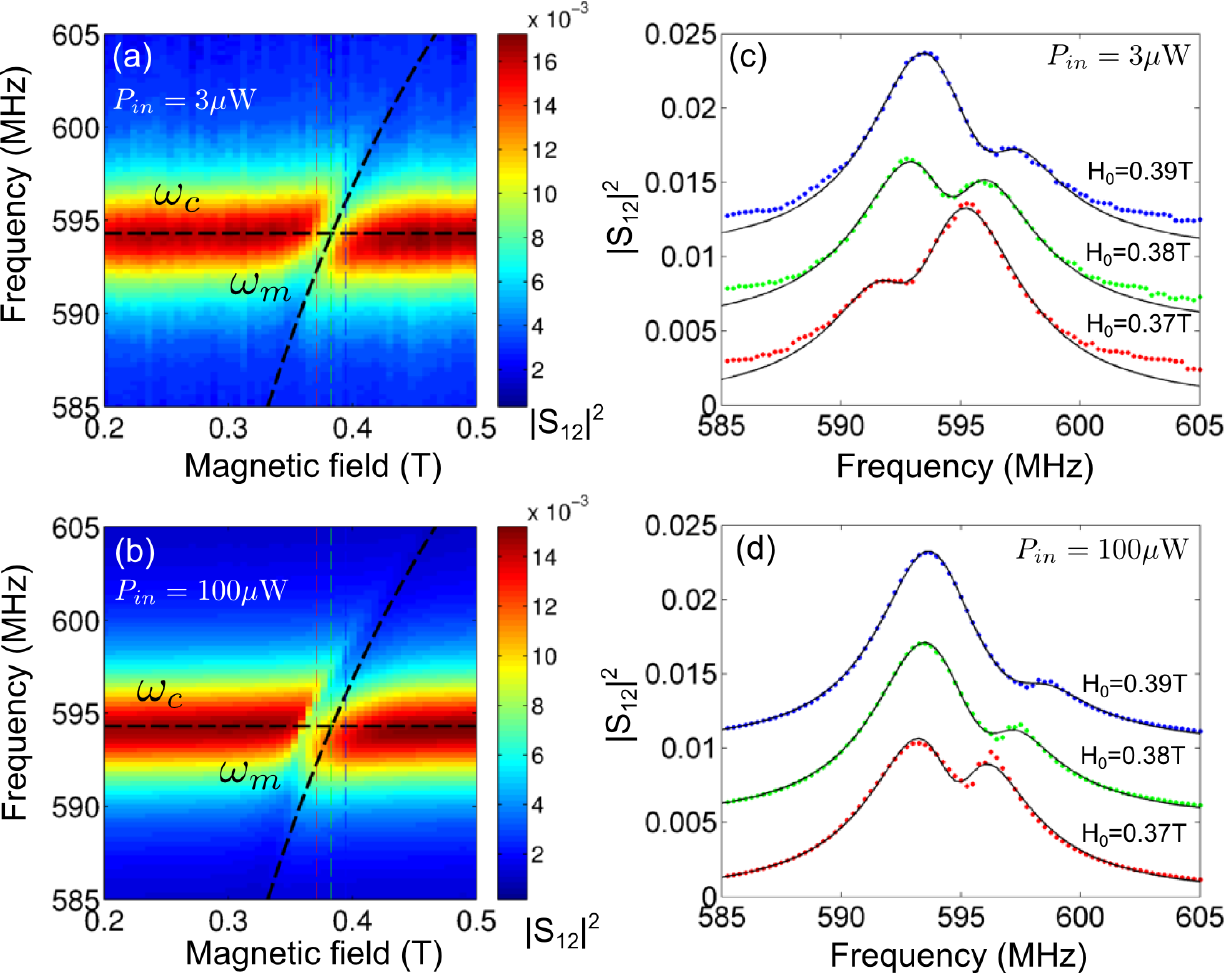}
\caption{\label{fig3} (color online) Transmission spectroscopy of the MnCO$_3$ sample. (a),(b) Transmission spectra as functions of the probe microwave frequency and the static magnetic field for input microwave power of $P_{in} = 3 \mu$W and $P_{in}=100\mu$W, respectively. The horizontal dashed line corresponds to the resonance frequency $\omega_c$ of the split-ring resonator. The dashed curve is the frequency $\omega_m$ of the hybridized nuclear magnon mode calculated by the Eq.~(\ref{eq1}). Normal-mode splitting is observed at approximately $H_0=0.38$\,T where the magnon mode $\omega_m$ and the microwave cavity mode $\omega_c$ would cross otherwise. (c),(d) Cross sections of the spectrum shown on panels (a) and (b), respectively, at static magnetic fields 0.37\,T, 0.38\,T, and 0.39\,T (shown by vertical dashed lines on panels (a),(b)). Dots are experimental data with vertical offset for clarity, and lines are the fitting curves.}
\end{figure*}

\paragraph{Normal-mode splitting.}
Figures~\ref{fig3}(a,b) show the microwave transmission $|S_{12}|^2=P_{out}/P_{in}$ as a function of probe microwave frequency $f=\omega/2\pi$ and applied magnetic field $H_0$ at the probe microwave power of $-25$\,dBm ($P_{in}\approx 3 \mu$W) and $-10$\,dBm ($P_{in}=100\mu$W), respectively. Resonances of the system are appeared as yellow or red regions. The horizontal dashed line corresponds to the resonance frequency $\omega_c$ of the split-ring resonator, while the dashed curve is the frequency $\omega_m$ of the hybridized nuclear magnon mode calculated by the Eq.~(\ref{eq1}) for $T_n = T = 1.15$\,K.  When the field approaches the value at which the magnon mode $\omega_m$ and the microwave cavity mode $\omega_c$ would cross, a noticeable avoided crossing is observed, indicating normal-mode splitting of the hybridized nuclear magnon mode of the MnCO$_3$ and the cavity mode.

With the increase of the probe microwave power, the avoided crossing was shifted slightly to the lower magnetic fields (see FIG.~\ref{fig3}(b)). At much higher levels of the probe microwave power ($0.3$\,mW\,$\leq P_{in} \leq 10$\,mW), the avoided crossing was shifted further to low fields, and was significantly distorted as compared with the data presented in FIG.~\ref{fig3}~\cite{note1}. The shift and the distortion of the splitting shape can be caused by nonlinear spin dynamics. Indeed, the nuclear magnetization $\mathbf{m}$ can significantly deviate from its equilibrium direction $\mathbf{m_0}$ at high microwave power, and formulas~(\ref{eq1})), derived under the assumption of small oscillations~\cite{deGennes1963}, are no longer applicable in this case. Recently, experimental and theoretical studies of nonlinear spin dynamics of MnCO$_3$ system have been performed, indicating the formation of the magnon Bose-Einstein condensate at high levels of microwave power~\cite{bunkov2012}.

We evaluate the coupling strength $g_m$ and the magnon and cavity linewidths by fitting the transmission spectrum $|S_{12}(\omega)|^2$ with an equation derived from a standard input-output formalism~\cite{huebl2013, tabuchi2014}:
\begin{equation}
|S_{12}|^2= \left| \frac{\sqrt{\kappa_1 \kappa_2}}{i(\omega-\omega_c)-\frac{\kappa_1+\kappa_2+\kappa_i}{2}+ \frac{g_m^2}{i(\omega-\omega_m)-\frac{\gamma_m}{2}}} \right|^2,
\label{eq3}
\end{equation}
where $\kappa_1$ and $\kappa_2$ are the external coupling rates to the resonator, $\kappa_i$ is the internal dissipation rate of the resonator, and $\gamma_m$ is the linewidth of the magnon mode. As shown by lines in FIG.~\ref{fig3}(c,d), the spectra agree well with the equation~(\ref{eq3}). 

The parameters obtained by the fitting are $g_m/2\pi \approx 1$\,MHz, $\gamma_m / 2\pi \approx 3$\,MHz, and the total resonator linewidth $\kappa/2\pi = (\kappa_1+\kappa_2+\kappa_i)/2\pi \approx 6$\,MHz. The cooperativity is calculated to be $C=4g_m^2/\kappa \gamma_m \approx 0.2$ which is close to the conditions of strong coupling regime defined by inequalities $g_m > \kappa, \gamma_m$ and $C > 1$.  

The obtained value of $\gamma_m$ is unexpectedly large. In the MnCO$_3$ system of coupled nuclear and electron spins at the temperature of $T \approx 1$\,K, the spin-lattice relaxation time of the hybridized nuclear-frequency mode is expected to be $T_1\approx 1$\,ms, while the spin-spin relaxation time should be of the order of $T_2 \approx 10$\,$\mu$s~\cite{bunkov1975}. Therefore, the observed linewidth of the magnon mode $\gamma_m/2\pi=3$\,MHz is probably determined by inhomogeneous broadening due to imperfections in the crystal (for example, crystal defects, and crystal inhomogeneities caused by mechanical stresses~\cite{shaltiel1966, tulin1969}). 

Next, we compare the experimentally determined coupling strength $g_m/2\pi \approx 1$\,MHz with theoretical estimations. The coupling strength between photons and the electron-spin component can be estimated as $g_e=(\gamma_e/2)\sqrt{\hbar \omega_c \mu_0/V_c}\sqrt{2NS} \times sin(\alpha)$. Here, the term $\sin(\alpha) \approx (H_{DM}+H_0)/2H_E$ is added to the standard formula (see, for example, \cite{huebl2013, zhang2014, tabuchi2014}) to take into account the canting angle $\alpha$ of the electron magnetic moment~\cite{pincus1960} which determines orientation of electron spins with respect to the microwave field. Inserting the values of the corresponding parameters, we obtain $g_e/2\pi \approx 3$\,MHz. Similarly, the strength of direct coupling between photons and nuclear spins is $g_n = (\gamma_n/2) \sqrt{\hbar \omega_c \mu_0/V_c}\sqrt{2 p_n N I}\times sin(\alpha)$. Since the nuclei are paramagnetic, the coupling strength between photons and nuclear spins depends on nuclear spin polarization $p_n$~\cite{eddins2014}. The net magnetization of the nuclear sublattice containing $N$ spins is $m=N \gamma_n^2 \hbar^2 I (I+1) H_n/(3 k_B T_n)$~\cite{abragam}, and, hence, spin polarization can be estimated as $p_n=\gamma_n \hbar (I+1) H_n/(3 k_B T_n)$. For our experimental conditions, spin polarization $p_n \approx 0.5 \%$   , and $g_n/2\pi \approx 100$\,Hz. 

By comparing $g_m$ with obtained values of $g_e$ and $g_n$, we conclude that the direct coupling between nuclear spins and photons is negligible, but the observed normal-mode splitting is mediated by coupling between photons and the electron-spin component of the hybridized magnon mode. Full description of the coupling between electron spins, nuclear spins, and photons is a challenging problem, but some estimations can be done by using the model of the enhanced microwave field $h_n$. As explained in the Introduction, influence of electron spin motion on the nuclei can be taken into account by considering the enhanced microwave field $h_n$ acting on the nuclei. The microwave field $h_n$ is enhanced by the factor $\eta \approx H_n/H_0 \approx 150$ relative to the applied field $h_1$, and is perpendicular to the equilibrium direction of the nuclear magnetization.  Hence, the actual coupling strength between nuclear spins and photons should be $G = (\gamma_n/2) \sqrt{\hbar \omega_c \mu_0/V_c}\sqrt{2 p_n N I}\times (H_n/H_0)$. Calculated value $G/2\pi \approx 1$\,MHz is in very good agreement with the measured value $g_m$.
 
We also note that $g_m$ is independent of the probe microwave power from $3\mu$W to $100\mu$W. This is consistent with results of the works~\cite{huebl2013,eddins2014, chiorescu2010}, since the number of photons in the resonator $N_\omega \approx (Q_c  P_{in} / \omega_c) / \hbar \omega_c \lesssim 10^{13}$ is much smaller than the number of spins $N$.  

\paragraph{Conclusions.}
For the first time, we observed the normal-mode splitting between the hybridized nuclear magnon mode in weak ferromagnetic MnCO$_3$ and the microwave mode of the split-ring resonator. According to our estimations, the coupling between the paramagnetic nuclear spin ensemble and the microwave cavity mode is mediated by the interaction between photons and the largely-detuned electron ferromagnetic-resonance mode which, in turn, is coupled to nuclear spins via the hyperfine interaction. The obtained cooperativity value $C=0.2$ indicates that our system is close to the conditions of strong coupling. By improving the resonator Q-factor and the quality of crystal, strong coupling can be achieved. The coupling strength can be also increased by cooling the system to millikelvins, and, hence, increasing the nuclear spin polarization. It would be also of great interest to consider possibility of multiple strong coupling under conditions of nuclear electron double resonance~\cite{shaltiel1964,tulin1969} when, in addition to the microwave radiation at the NMR frequency, microwave radiation at the electron FMR frequency is applied (see FIG.~\ref{fig1}(b)). We suppose that similar strong coupling phenomena can be realized in other systems with the strong hyperfine interaction (CsMnF$_3$, MnF$_2$, FeBO$_3$, CoCO$_3$, and other systems~\cite{borovik-romanov1984}).

\begin{acknowledgments}
The work was supported by an internal grant from Okinawa Institute of Science and Technology (OIST) Graduate University, and JSPS KAKENHI (Grant No. 26400340). Yu.M.B. was supported as Visiting Researcher by Okinawa Institute of Science and Technology (OIST) Graduate University.
\end{acknowledgments}

\bibliography{couplingMnCO3_ref.bib}

\begin{thebibliography}{39}
\bibitem [{not()}]{note1}%
  \BibitemOpen
  \href@noop {} {}\bibinfo {note} {Results obtained at high values of the probe
  microwave power will be published elsewhere.}\BibitemShut {Stop}%
\end{thebibliography}
\end{document}